\documentclass[a4paper,11pt]{article}
\usepackage{pos}

\title{Hard and electromagnetic probes:\\
plans for future measurements at the CERN SPS}

\author*[a]{Enrico Scomparin}

\affiliation[a]{Istituto Nazionale di Fisica Nucleare,\\
  Via Giuria 1, Torino, Italy}

\emailAdd{scomparin@to.infn.it}

\abstract{The CERN SuperProtoSynchrotron (SPS) represents an ideal facility for fixed-target heavy-ion experiments exploring the phase diagram of strongly interacting matter in the region $200\le\mu_{\rm B}\le500$ MeV. It can deliver high-intensity beams ($>10^6$ Pb/s), allowing a study of rare probes of the Quark-Gluon Plasma, including electromagnetic and hard processes. The NA61/SHINE experiment is currently active and plans to perform a first direct measurement of open charm production in Pb--Pb collisions at top SPS energy and possibly at lower energies. The project of a new experiment, NA60+, based on a muon spectrometer coupled to a vertex spectrometer is currently being developed, for the study of dimuon and heavy quark production, and a Letter of Intent was recently submitted. In this contribution the physics motivation for the studies of rare probes, the existing and planned experimental set-ups and their expected physics performance will be discussed. }

\FullConference{
 26-31 March 2023 \\
 Aschaffenburg, Germany\\}

\begin{document}
\maketitle

\section{Introduction}

Studies of electromagnetic and hard probes have represented one of the most important sources of information on the early stages of the strongly interacting partonic system created in ultra-relativistic heavy-ion collisions. From 1986 to 2003, historic first results in this domain were obtained by experiments taking data at the CERN SPS. Among those, one can mention the first evidence for a modification of the $\rho$-meson spectral function by the CERES/NA45 experiment~\cite{CERES:2005uih}, the discovery of the anomalous suppression of the J/$\psi$ by NA50~\cite{Alessandro:2004ap} and the first direct measurement of a temperature exceeding $T_{\rm c}$, via thermal dimuons by NA60~\cite{Arnaldi:2008er}.
However, all of these measurements were performed at top SPS energy ($\sqrt{s_{\rm NN}} = 17.3$ GeV), with the extension to lower energies made difficult by the quickly vanishing production cross sections. Among the hard probes of QGP, open charm production was also marginally explored, with an indirect result from NA60 obtained in the study of the dimuon spectrum in the region around $m_{\rm \mu\mu}\sim 2$ GeV/$c^2$~\cite{Arnaldi:2008er} and an upper limit on D$^0$ production~\cite{Alt:2005zu} from NA49.

An extension of all these measurements to lower energies presents a great physics interest. Concerning thermal dileptons, their emission takes place all along the collision history but the yield is dominated by the high-temperature phase. The $T_{\rm slope}$ parameter of the spectral shape of the dilepton invariant mass spectrum can be considered as a space-time average of the thermal temperature $T$ over the  fireball evolution. Measurements carried out by the HADES~\cite{HADES:2019auv} and NA60~\cite{Arnaldi:2008er} experiments, respectively in Au--Au collisions at $\sqrt{s_{\rm NN}} = 2.42$ GeV and In--In at $\sqrt{s_{\rm NN}} = 17.3$ GeV, allowed experimental estimates of the medium temperature. Values increasing from $T_{\rm slope} = 71.8\pm2.1$ MeV to $205 \pm 12$ MeV were obtained, showing that the crossing of the pseudo-critical temperature should be reached in the collision energy interval defined by the two existing measurements. A precise measurement of the dilepton spectrum below top SPS energy is therefore quite interesting and a precision study of its $\sqrt{s_{\rm NN}}$-dependence might also be sensitive to the presence of a first order phase transition~\cite{Rapp:2014hha}. Measurements of lepton pairs also allow a study of modifications of the spectral function of the $\rho_0$ and of its chiral partner $a_{\rm 1}$, related to chiral symmetry restoration. The latter resonance is not directly coupled to the dilepton channel, but its modification leads to an enhancement of the continuum in the region $1<m_{\rm ll}<1.4$ GeV/c$^2$~\cite{Rapp:2014hha}.

Concerning hard probes of the QGP, quarkonium production can also be investigated by a dilepton measurement.  Studying the J/$\psi$ suppression, and possibly that of higher-mass states ($\psi({\rm 2S})$, $\chi_{\rm c}$) is of utmost importance, seen its relation with deconfinement, studied in great detail from top SPS~\cite{Arnaldi:2007zz} to RHIC~\cite{Adare:2011yf} and LHC energy~\cite{ALICE:2023gco}. A superposition of a suppression mechanism, driven by color screening and gluodissociation processes, and regeneration at hadronization is able to explain the wealth of data at collider energies. Extending measurements towards lower energies would allow the identification of a threshold for the suppression, which is the dominant mechanism at low energy, and its correlation with the temperature measured via thermal dileptons. In addition to hidden charm, open charm production represents a crucial measurement to get information on the transport coefficients of the medium, as well as on its hadronization mechanisms~\cite{Prino:2016cni}. At low energy, the system spends comparatively more time in the hadronic phase than in the QGP, and it is not clear to which degree heavy quark would thermalize in the medium. Accurate results on $R_{\rm AA}$ and $v_{\rm 2}$ of charm hadrons would test an environment very different from that accessible to collider experiments and represent a stringent testbench for models. Furthermore, separate measurements of charmed baryon over meson and strange over nonstrange hadron ratios would allow a study of coalescence effects around the threshold for QGP production. 
Clearly, accurate measurements of open charm hadrons can only be performed by accessing their hadronic decays, requiring the capability of tracking/identifying charged particles in a high-multiplicity environment. 

The physics program sketched in these paragraphs can be addressed by a new SPS experiment, NA60+~\cite{NA60:2022sze}, based on a muon spectrometer coupled to a vertex spectrometer, and inspired to the former NA60 experiment. Specific items, and in particular a first direct measurement of open charm production, can be addressed by the existing NA61/SHINE~\cite{NA61:2014lfx} experiment at the same facility.
In the next sections details will be given on the foreseen measurements by NA61 and on the progress of the design and R\&D phase of the NA60+ experiment. For the latter, the current estimates of the physics performance and the planning of the data taking phase will be given.

\section{Open charm measurements at NA61/SHINE}

\begin{figure}[htb]
\centering{
\includegraphics[width=6.35cm]{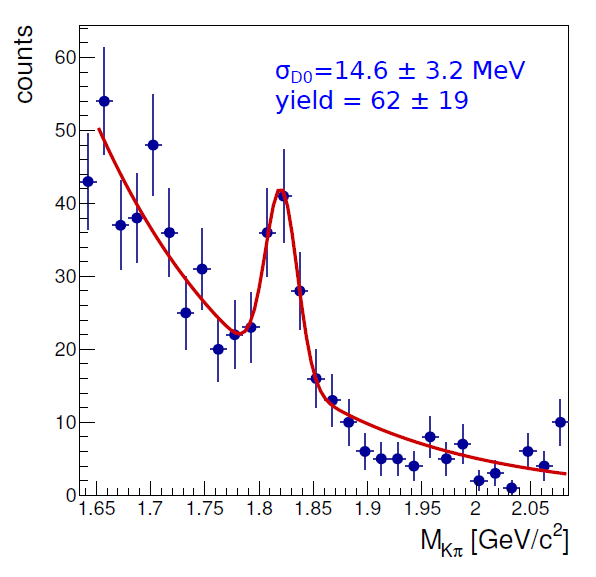}}
\caption{Invariant mass distribution of D$^0$ + $\bar{\rm D}^0$ candidates in central Pb--Pb collisions, from NA61/SHINE, after background suppression cuts~\cite{Brylinski:2022ewb}.}
\label{fig:na61D}
\end{figure}

NA61/SHINE, active since 2009, is a multi-purpose experiment investigating hadron production in nuclear collisions. It has performed systematic studies with various ion species at different energies, investigating signals of the onset of deconfinement. Among the main physics goals for 2022-2025, the experiment is planning a measurement of open charm, also thanks to an upgrade of the vertex detector to stand a 1 kHz interaction rate.
The expected statistics is $5\cdot 10^8$ minimum bias events at $E_{\rm lab}=150 A$ GeV and $2.5\cdot 10^8$ at $E_{\rm lab}=40 A$ GeV. The data should provide, at high energy, the D-meson rapidity and transverse momentum spectra for central collisions (tens of thousands reconstructed decays) and establish the centrality dependence of mean D-meson yield in the acceptance (thousands of reconstructed decays). Together with the low-energy data, the first measurement of the collision energy dependence of the mean yield should become possible. A low-statistics invariant mass spectrum is shown in Fig.~\ref{fig:na61D}.

\section{The NA60+ experiment}

\begin{figure}[htb]
\centering{
\includegraphics[width=14cm]{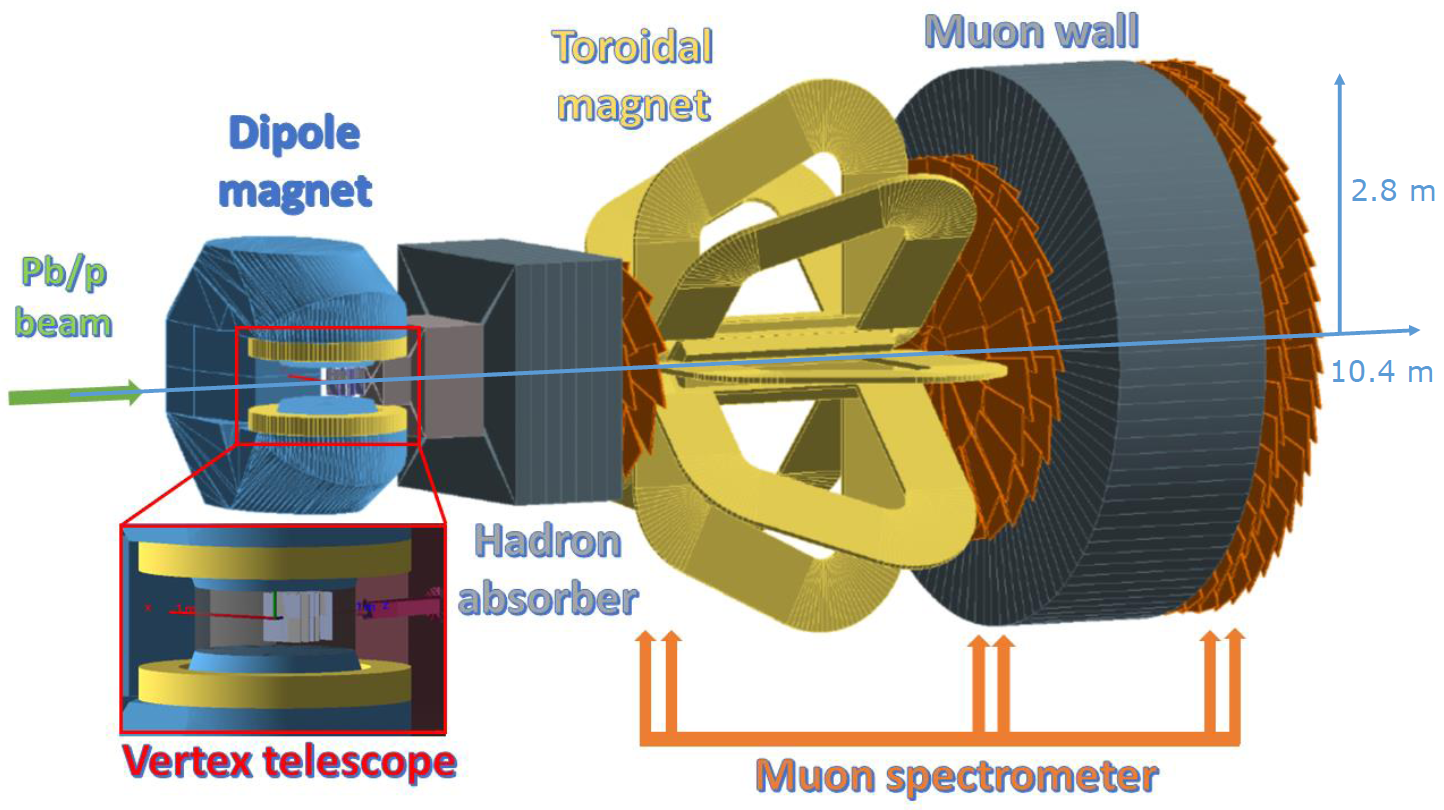}}
\caption{The foreseen set-up of the NA60+ experiment at the CERN SPS.}
\label{fig:NA60plussetup}
\end{figure}

A sketch of the NA60+ experiment can be seen in Fig.~\ref{fig:NA60plussetup}. A muon spectrometer is separated from the target region by means of a thick hadron absorber, made of BeO and graphite that filters out the hadrons. Candidate muons are tracked by means of six tracking stations. The two upstream stations are followed by a large-aperture toroidal magnet and by two further tracking stations. A ``muon wall'' acts as a final filter to remove hadrons escaping the absorber as well as low-momentum muons, and is followed by the two final tracking stations. In the target region a system of five closely spaced targets is followed by a vertex spectrometer which includes from five up to ten tracking MAPS planes. The vertex spectrometer is embedded in the gap of a dipole magnet. By matching muon tracks in the muon and vertex spectrometers, in the coordinate and momentum space, a  measurement of the muon kinematics not affected by multiple scattering and energy loss in the hadron absorber becomes possible. The muon spectrometer is mounted on a system of rails that allows keeping a constant rapidity coverage when changing the collision energy. When moving the spectrometer downstream along the beam line (high energy collisions) the thickness of the absorber will be increased by adding further graphite elements.

\begin{figure}[htb]
\centerline{%
\includegraphics[width=6.35cm]{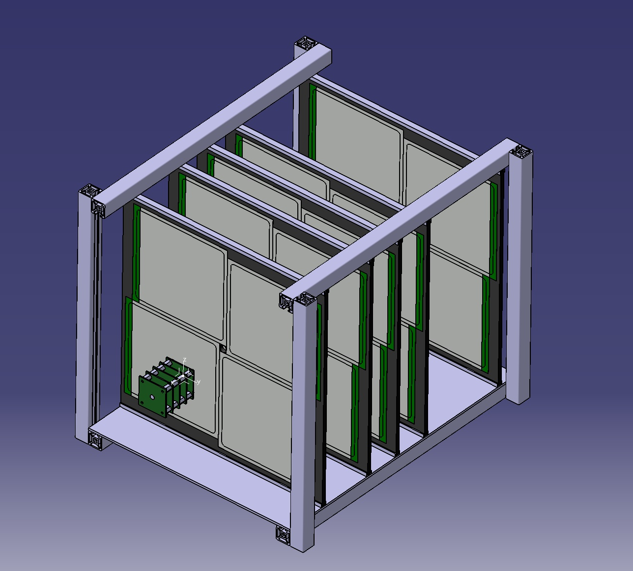}
\includegraphics[width=5.65cm]{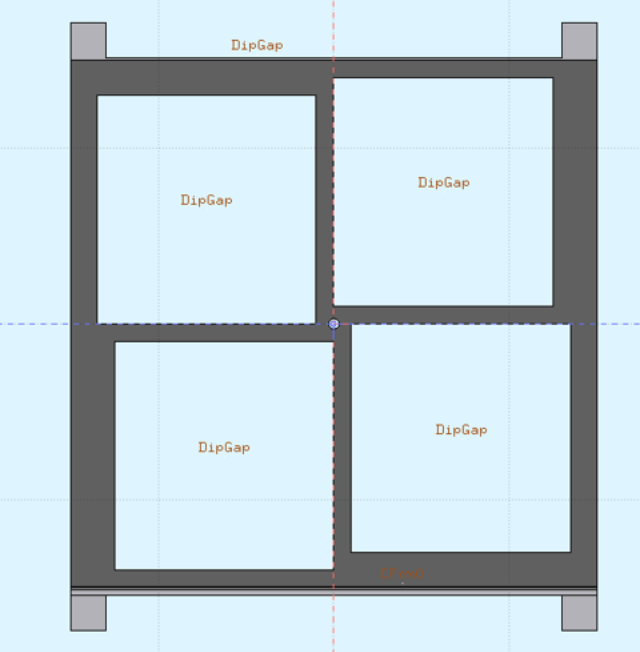}}
\caption{Possible layout of the target system and the vertex spectrometer (left); sketch of the mechanical support of a MAPS plane (right).}
\label{fig:target}
\end{figure}

In Fig.~\ref{fig:target} (left) a possible arrangement of the vertex spectrometer is shown. Each plane contains four large MAPS sensors ($15\times15$ cm$^2$ each). The R\&D phase is in progress in the frame of a common development of the ALICE (ITS3 project~\cite{Musa:2703140}) and NA60+ collaborations. For NA60+ each sensor is based on 25 mm long units, replicated several times through a stitching procedure. For each stitched long sensor, the control logic to steer the priority encoders, the interfaces for the configuration of the chip and serial data transmitters are all located at the periphery of the sensor, outside the detector acceptance. The thickness of the sensor will be $<0.1$\% radiation length and the spatial resolution will be $\le 5$ $\mu$m. From the mechanical point of view the sensors will be mounted on a light frame, as shown in Fig.~\ref{fig:target}(right). Studies of cooling options show that a mixed fluid (0.5 l/s at T=18 $^{\rm o}$C) + air ($\sim$ 1 m/s) system should allow an efficient control of the temperature over all the surface of the detection plane. A test set-up is being built, with thin graphite frames, in order to perform tests of positioning and gluing of sensors (dummy Si sensors are used for these studies). It will also be possible to check the chosen cooling option using a flex circuit with resistor arrays, glued on the sensor, to simulate the expected power dissipation. Flex cables for wire-bonding tests and 
high-speed data transmission will also be designed.

\begin{figure}[h]
\centerline{%
\includegraphics[width=6cm]{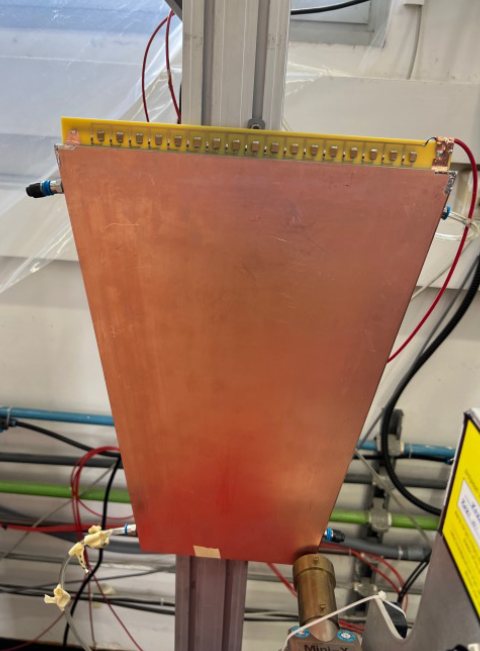}
\includegraphics[width=6cm]{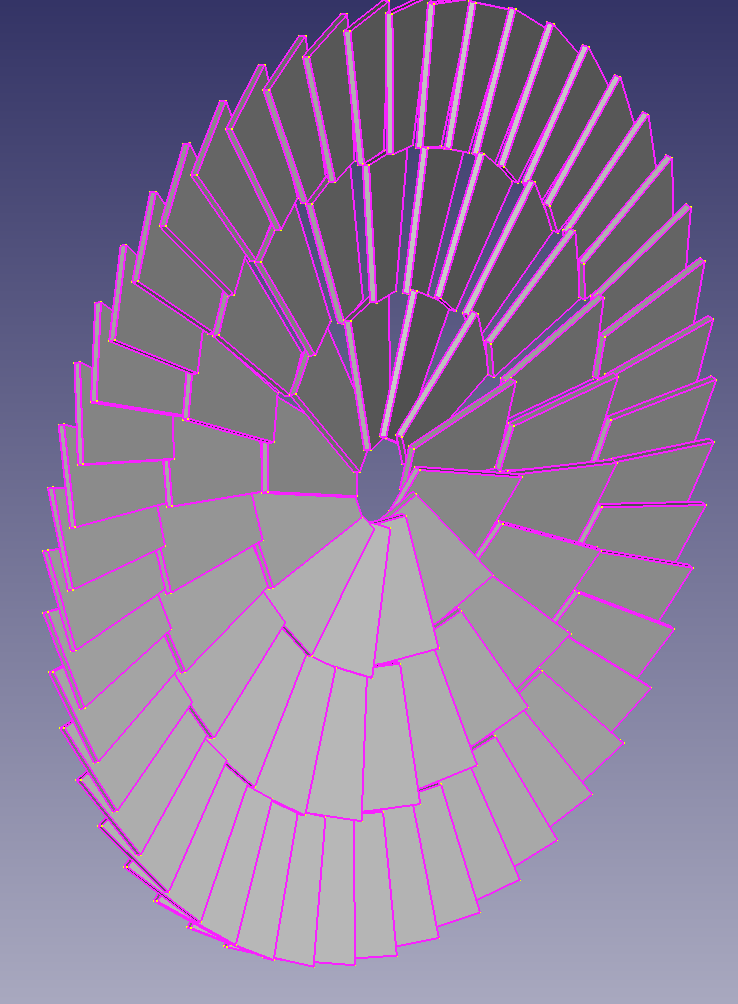}}
\caption{First prototype of a MWPC unit for the NA60+ muon spectrometer (left); layout of the largest muon tracking stations, downstream of the Muon wall (right).}
\label{fig:muonspectro}
\end{figure}

For the muon tracking, due to the relatively low rates after the hadron absorber ($\sim 2$ kHz, at the expected Pb beam rate of $10^6$ s$^{-1}$) GEMs or MWPCs options are considered. For the latter, two prototypes were built and tested in the laboratory with cosmic rays. They will be exposed to an SPS test beam at the end of 2023. The tracking stations will be built by replicating the prototype detector in such a way to obtain the desired geometry. The two upstream stations will include 12 trapezoid detectors each, while those downstream of the magnet and the muon wall will have 36 and 84 detectors respectively. Figure~\ref{fig:muonspectro} shows an MWPC prototype unit (left) and the foreseen arrangement for the largest stations (right). The GEM option is also under study, based on the design of a triple GEM detector for the MOLLER experiment at JLab~\cite{MOLLER:2014iki}. Also this detector will be tested on an SPS beam this year.

\begin{figure}[h!]
\centerline{%
\includegraphics[width=9cm]{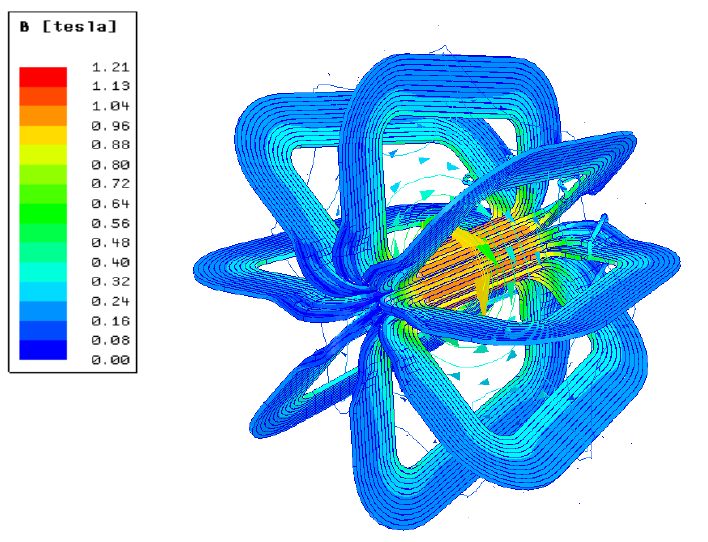}
\includegraphics[width=6cm]{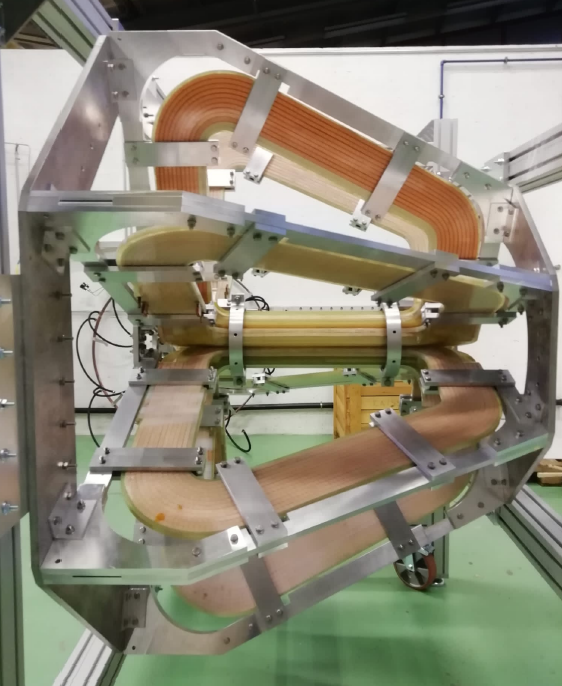}}
\caption{Magnetic field calculations for the NA60+ toroidal magnet (left); the scale 1:5 demonstrator with its support mechanics (right).}
\label{fig:magnet}
\end{figure}

The toroidal magnet represents a major element of the set-up. A warm pulsed magnet is foreseen, producing 0.5 T of magnetic field magnitude over a volume of 120 m$^3$.
The toroid is made from eight sectors, each coil has 12 turns and the conductor has a square copper section with 50 mm side length and a circular cooling channel in the centre. The current is 190 kA, and the total power is $\sim$3 MW. 
A demonstrator (scale 1:5) was constructed and tested, allowing a cross-check of various aspects of the design~\cite{ToroidalPrototypeNa60plus}. Measurements of the magnetic field in the prototype were found to be in agreement with simulations within 3\%. In Fig.~\ref{fig:magnet} (left) the results of the field calculations for the toroidal magnet are reported, while on the right panel a photo of the demonstrator can be seen. Finally, for the dipole magnet the plan is to use the MEP48 magnet, currently stored at CERN. It can deliver a 1.5 T field, over a 40 cm gap. 

After integration studies, the PPE138 hall on the H8 beam line of the SPS was chosen as installation site for the NA60+ set-up~\cite{Gerbershagen:2022zbq}. Radio-protection calculations have led to the design of an iron/concrete shielding, to keep the dose levels below 3 $\mu$Sv/h externally to the experimental hall for the intended 10$^6$ s$^{-1}$ Pb beam intensity. Specific beam optics studies have shown that a sub-millimetric beam can be delivered to the experiment from $E_{\rm lab}=30$ to 150 AGeV. A first check of these optics was carried out, making use of a telescope of ALPIDE Si sensors, in the fall of 2022 at $E=150$ GeV. Values as small as $\sigma_{\rm x}=280$ $\mu$m and $\sigma_{\rm y}=250$ $\mu$m were measured. 

\begin{figure}[htb]
\centerline{%
\includegraphics[width=9cm]{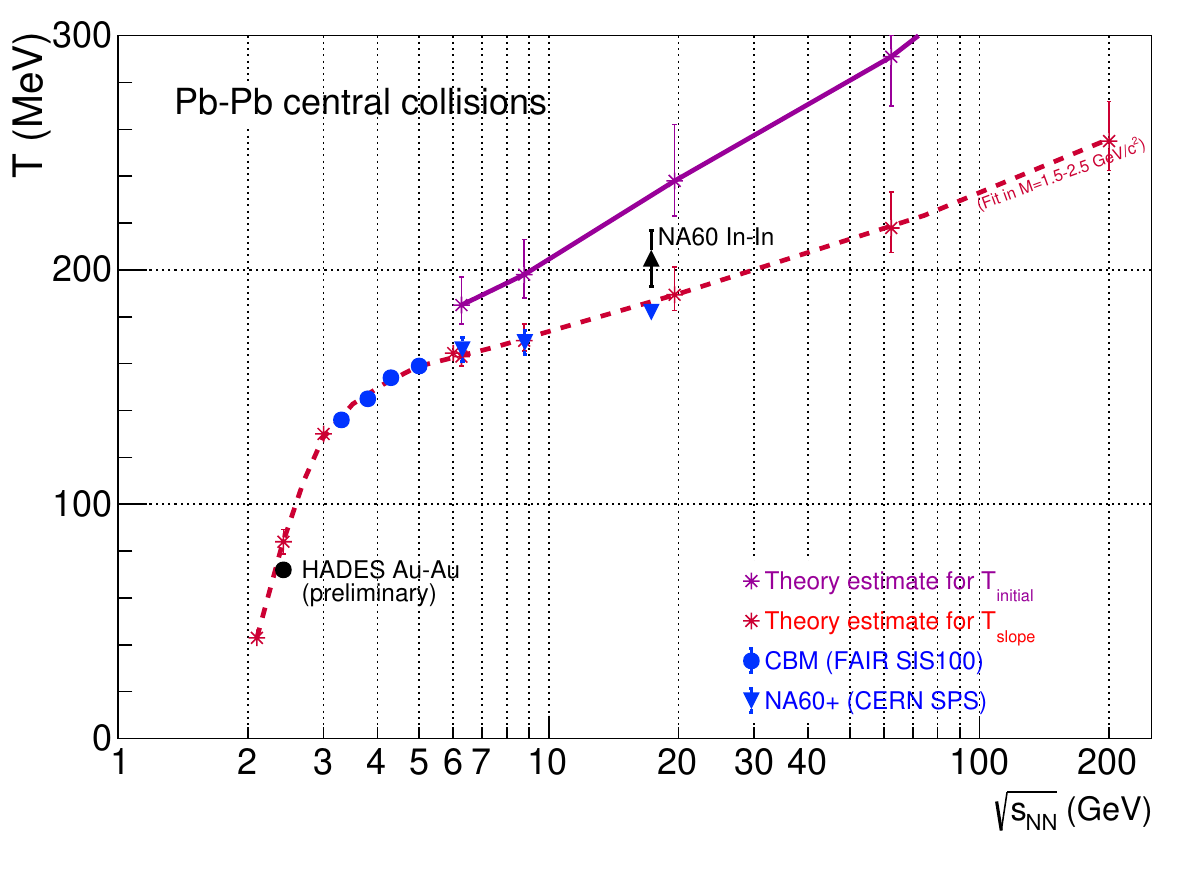}
\includegraphics[width=7cm]{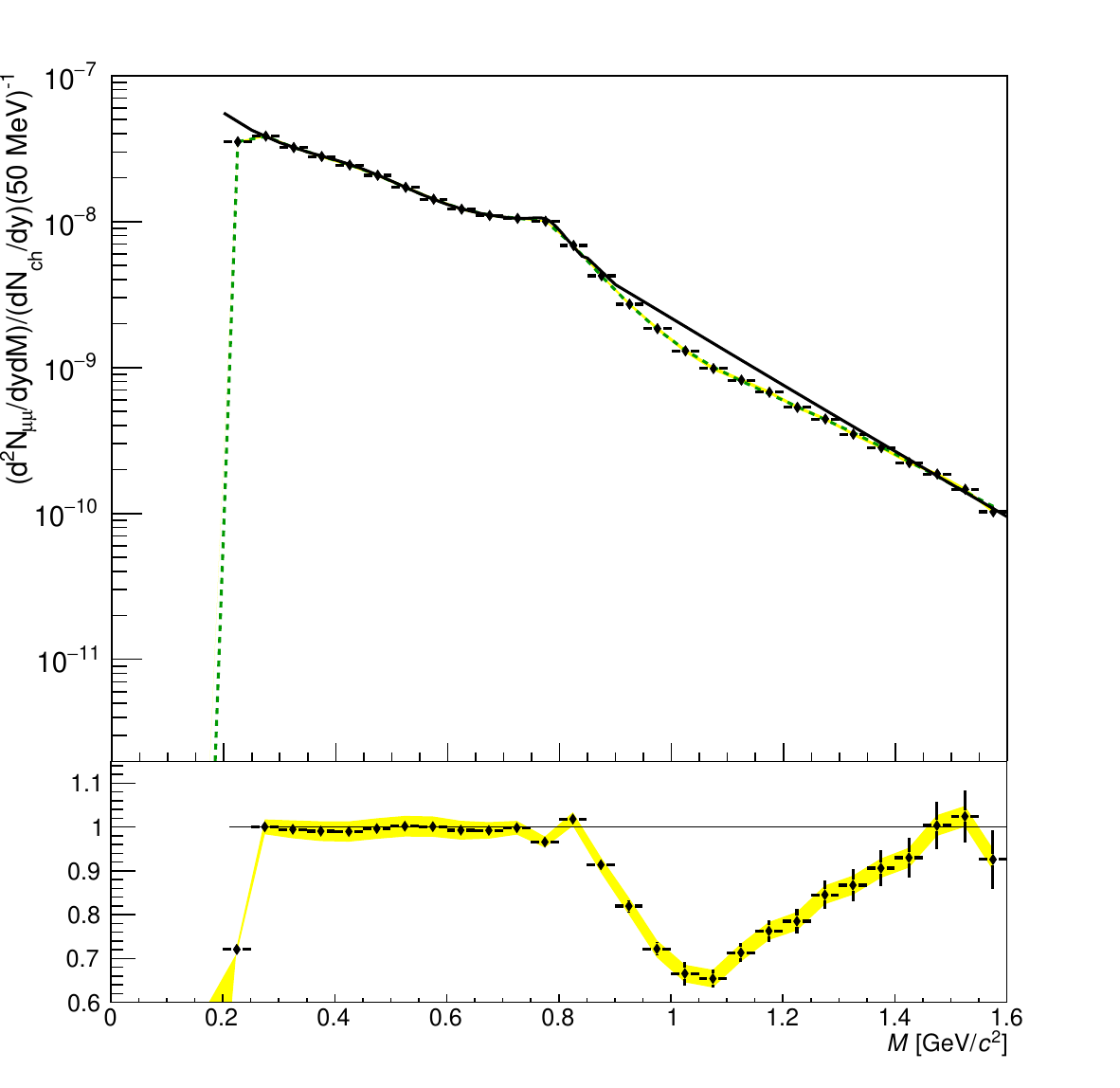}}
\caption{The energy dependence of the $T_{\rm slope}$ parameter (left), the blue triangles show the expected NA60+ performance; comparison of the thermal dimuon spectrum at $\sqrt{s_{\rm NN}}=8.8$ GeV in case of no chiral mixing and full chiral mixing (black line) (right). The ratio between the two spectra is also shown.}
\label{Fig:physics}
\end{figure}

Accurate studies of the expected physics performance of the NA60+ experiment were carried out. They are extensively reported in~\cite{NA60:2022sze} and a selection of the results is reported here. With the expected Pb beam intensity, 10$^{11}$ minimum bias collision events should be collected in the typical 1-month period dedicated to heavy-ion running at CERN each year. 
For the measurement of the $T_{\rm slope}$ parameter, in the mass region $1.5<m_{\mu\mu}<2.5$ GeV/$c^2$, uncertainties as low as $\sim 3\%$ can be reached. The result on the measurement of the energy dependence of $T_{\rm slope}$ is  shown in Fig.~\ref{Fig:physics} (left) together with existing measurements from HADES and NA60 and corresponding studies for the CBM experiment at GSI, which explores a lower, complementary energy range. It will be possible to explore the region close to the pseudocritical temperature and detect possible signals of a first-order phase transition that can be expected at these energies. The expected effect of a full chiral mixing on the dilepton spectrum is shown in Fig.~\ref{Fig:physics} (right). An increase of the dimuon yield by up to $\sim$30\% in the region around $m_{\mu\mu}=1$ GeV/$c^2$ is visible and should be detectable within the expected uncertainties.

\begin{figure}[htb]
\centerline{%
\includegraphics[width=8cm]{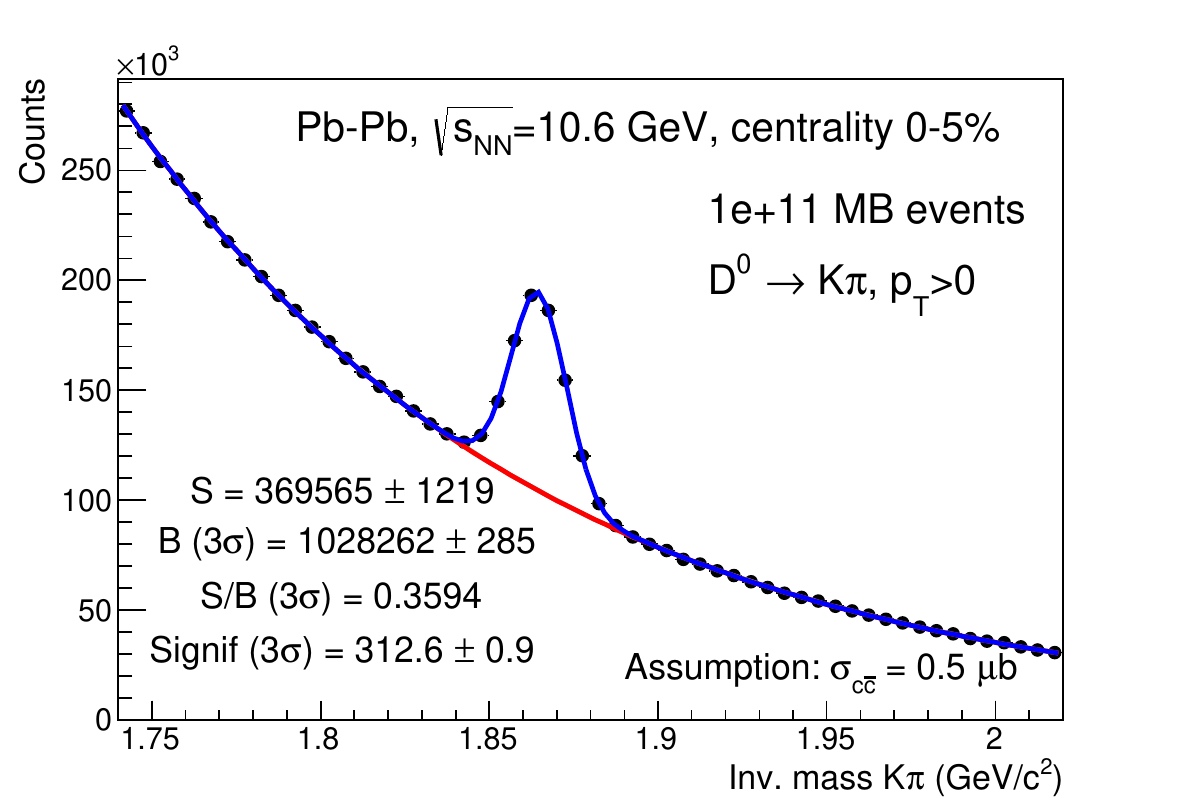}
\includegraphics[width=8cm]{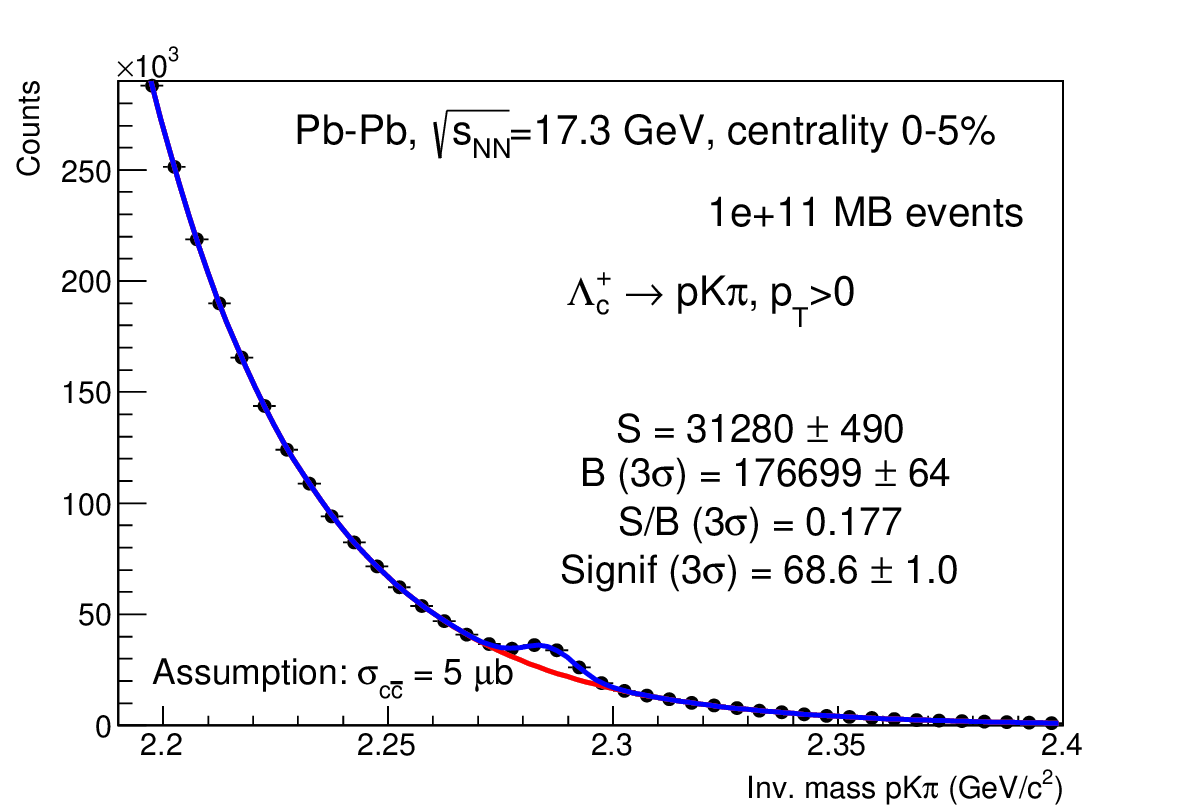}
}
\caption{Invariant mass distribution of candidates for the $D^{\rm 0}\rightarrow{\rm K}\pi$ decay at $E=60$ AGeV (left) and the $\Lambda_{\rm c}\rightarrow {\rm pK}\pi$ decay at $E= 158$ AGeV (right).}
\label{Fig:charm}
\end{figure}

Open charm measurements will be performed starting from the tracks reconstructed in the vertex spectrometer. Charm hadron candidates are built by combining pairs or triplets of tracks with the proper charge signs. The huge combinatorial background can be reduced via geometrical selections and exploiting the fact that the mean proper decay lengths $c\tau$ are of about 60–310 $\mu$m, and therefore their decay vertices
are typically displaced by a few hundred $\mu$m from the primary interaction vertex. 
In Fig.~\ref{Fig:charm} the expected signals from the ${\rm D}^{\rm 0}\rightarrow{\rm K}\pi$ decay, for $E= 60$ AGeV incident energy and from the 3-body decay $\Lambda_{\rm c}\rightarrow {\rm pK}\pi$, for $E= 158$ AGeV are shown. Measurements of the energy dependence of ${\rm D}^{\rm 0}$ production, as well as of its nuclear modification factor and $v_{\rm 2}$ at top SPS energy are within reach. Also studies of charmed baryon/meson ratios, sensitive to the hadronization mechanisms, will be possible. 

\begin{figure}[htb]
\centerline{%
\includegraphics[width=9.5cm]{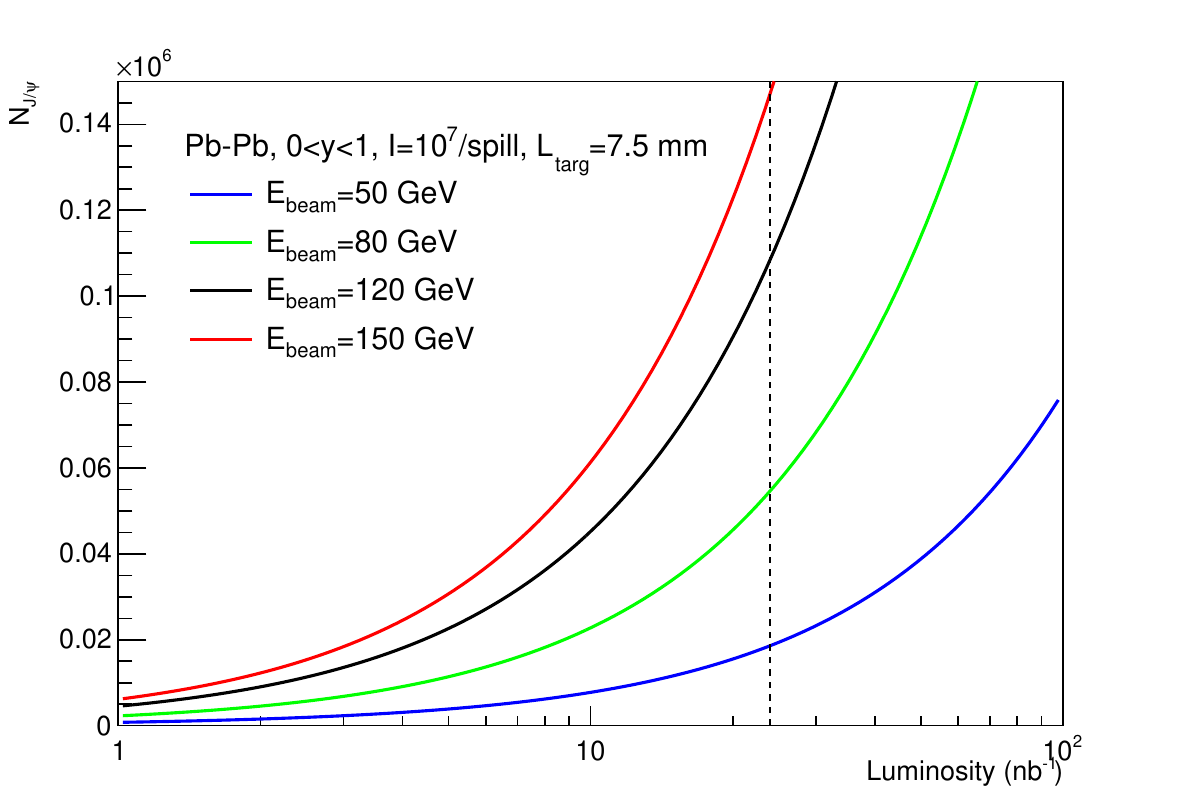}
\includegraphics[width=6.6cm]{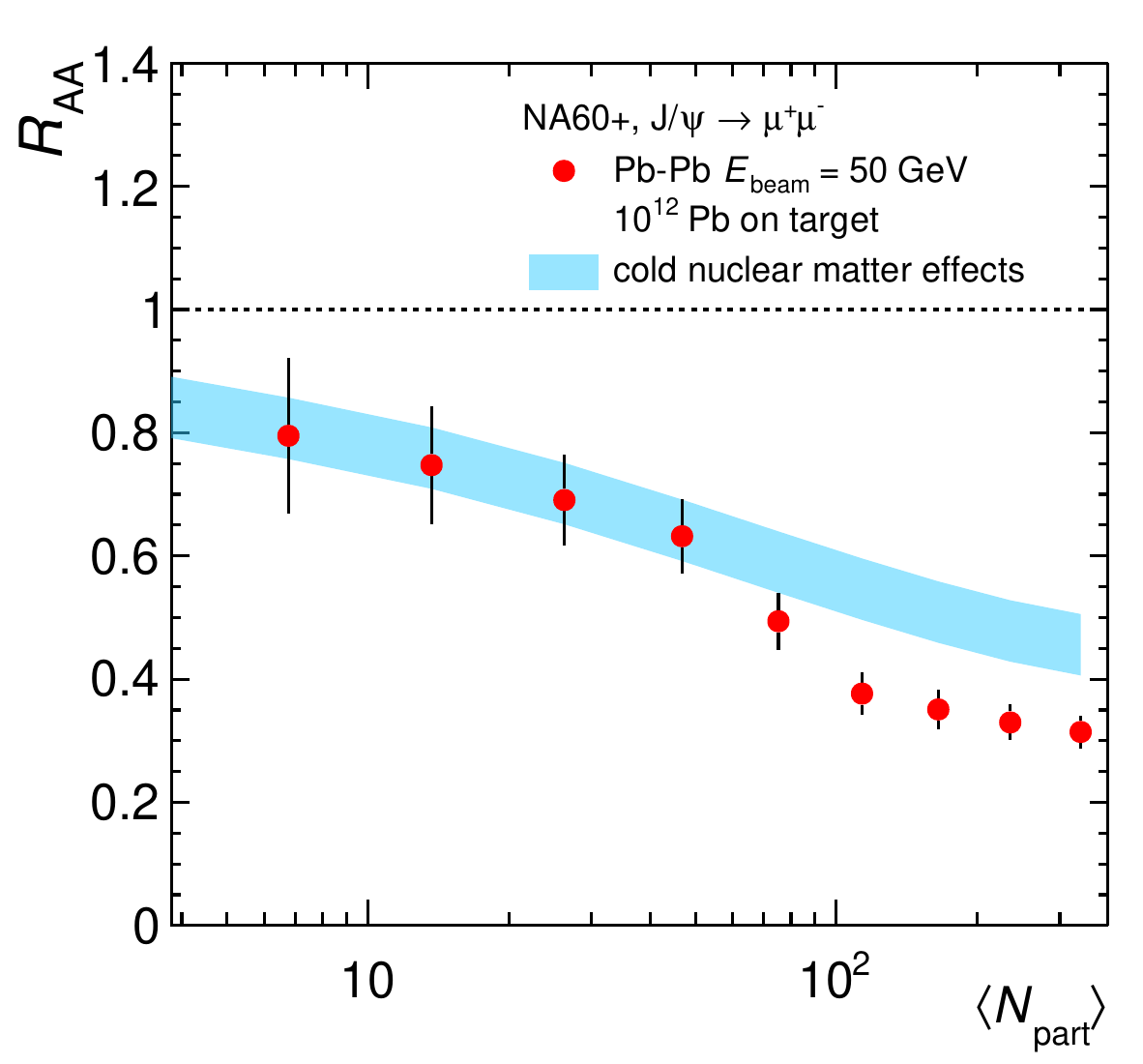}
}
\caption{The expected number of detected ${\rm J}/\psi\rightarrow\mu\mu$ decays, as a function of the integrated luminosity (left); the centrality dependence of the nuclear modification factor for the J/$\psi$ in Pb--Pb collisions at $E_{\rm lab}= 50$ AGeV (right), superimposed to the corresponding values for cold nuclear matter effects, extrapolated from the results of p--A collisions (see text for the assumptions).}
\label{fig:quarkonium}
\end{figure}

Measurements of hidden charm production will  also be performed. The expected statistics range from $10^4$ to $10^5$ J/$\psi$ from low to high SPS energy, allowing a precise study of the suppression signal, that is expected to vanish at low energy. Corresponding measurements in p-A collisions will also be carried out. The latter are crucial in order to establish the contribution of cold nuclear matter effects that is expected to become stronger at low energy. In Fig.~\ref{fig:quarkonium} (left) the number of ${\rm J}/\psi\rightarrow\mu\mu$ decays that can be detected as a function of the integrated luminosity, at various energies, is shown. With one month of data taking for Pb--Pb collisions one has $L_{\rm int}\sim 25$ nb$^{-1}$. In Fig.~\ref{fig:quarkonium} (right) a possible centrality dependence of the J/$\psi$ nuclear modification factor, at $E_{\rm lab}= 50$ AGeV, is shown. The cold nuclear matter reference is obtained by means of an extrapolation, using the Glauber model, of results on J/$\psi$ production in p--A collisions, obtained with $5\cdot 10^{13}$ protons on a target system made of various nuclear species. As an hypothesis, a 20\% suppression effect in Pb--Pb, beyond cold nuclear matter effects and starting at $N_{\rm part}\sim 100$, has been assumed. Although with larger uncertainties, corresponding measurements for the $\psi(2S)$ meson should be feasible down to $E_{\rm lab}\sim$ 100 GeV.

\begin{figure}[htb]
\centerline{%
\includegraphics[width=7.3cm]{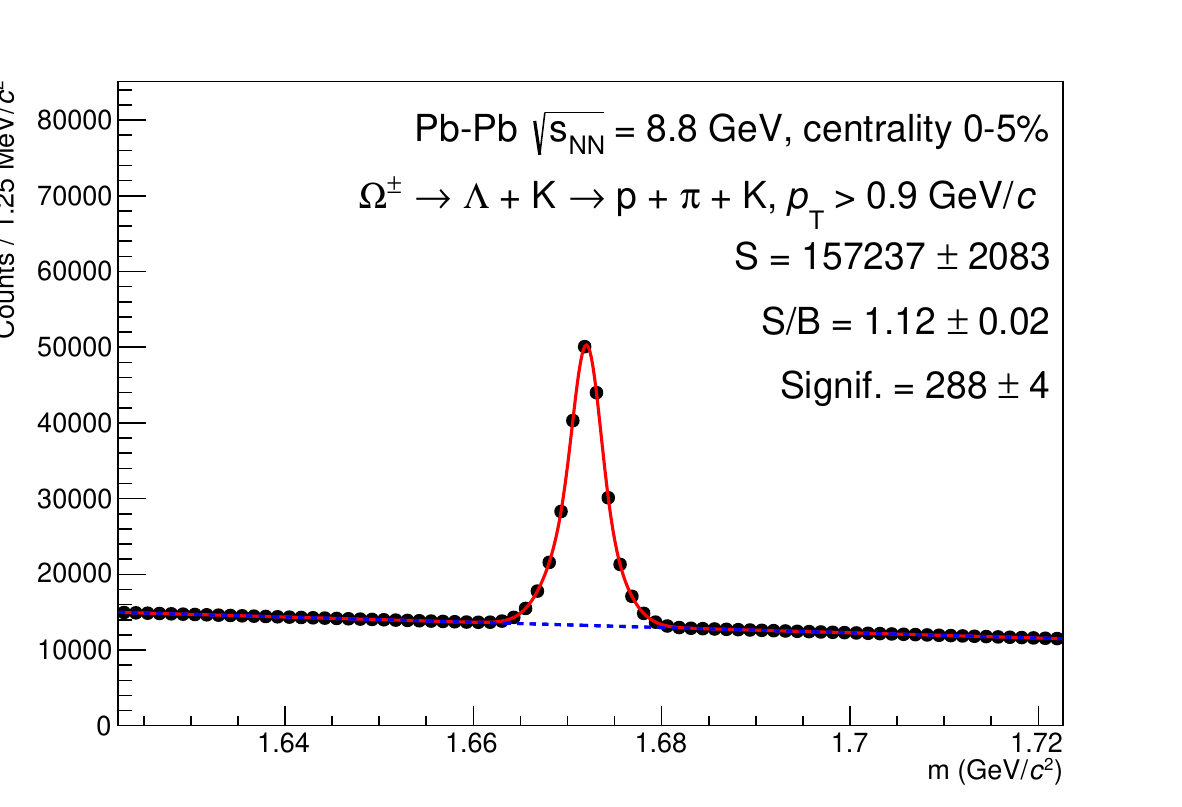}
\includegraphics[width=8.3cm]{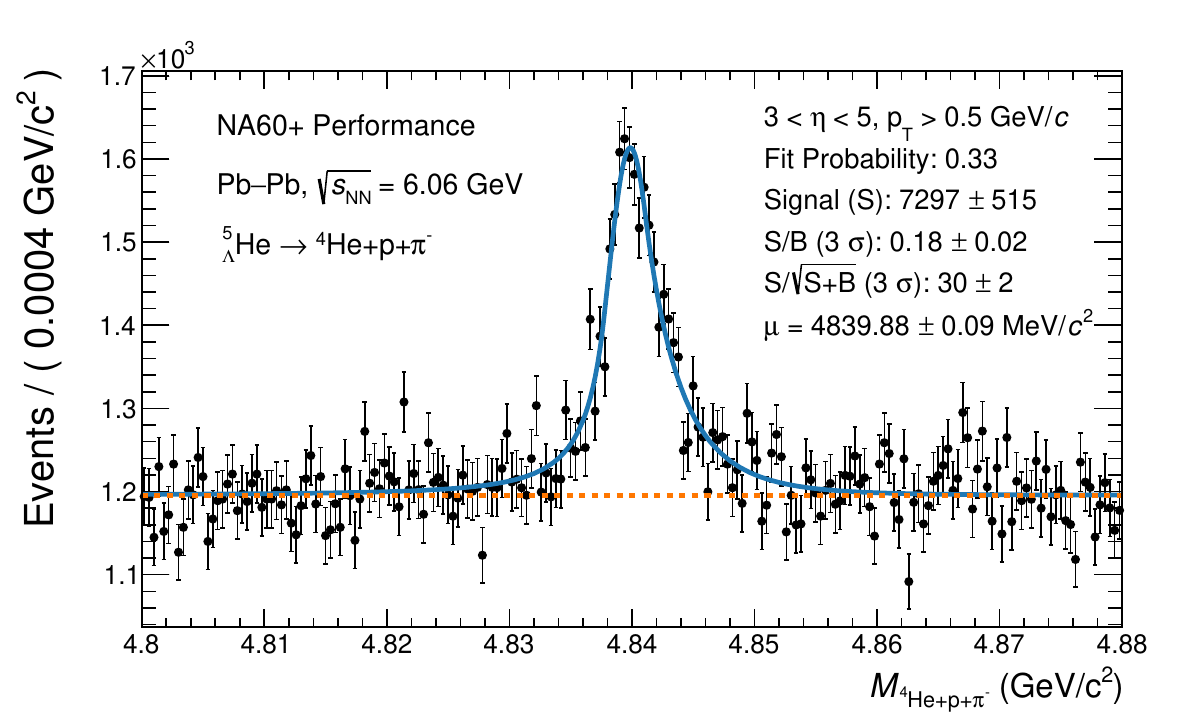}
}
\caption{Invariant mass spectrum from the $\Omega^{-}\rightarrow\Lambda^0 {\rm K}^-$ (+ charge conjugate) decay, at $E_{\rm lab}=40$ AGeV (left), and for the $^5_\Lambda {\rm He}$ hypernucleus (right).}
\label{fig:strangeness}
\end{figure}

Finally, in Fig.~\ref{fig:strangeness} (left), the invariant mass spectrum from the $\Omega^{-}\rightarrow\Lambda^0 {\rm K}^-$ decay (+ charge conjugate), is shown, as an example of the NA60+ capabilities for the measurement of hyperons in nuclear collisions. A realistic background was simulated, starting from the parametrization of hadron production from NA49 results, topological selections were then applied and Boosted Decision Trees were employed to enhance the significance of the signals. In Fig.~\ref{fig:strangeness} (right) an invariant mass spectrum for the decay $^5_\Lambda {\rm He}\rightarrow ^4 {\rm He}\,{\rm p}\,\pi^-$ is shown, demonstrating the possibility for NA60+ to detect hypernuclei. The separation of heavily ionising particles from ordinary hadrons, essential for the identification of the decay nucleus, is performed by means of a study of the cluster size associated with the track in the vertex spectrometer.

In summary, the NA60+ experiment should be able to access for the first time e.m. and hard probes from low to top SPS energy, via dimuon measurements and reconstruction of hadronic decays of open charm hadrons. Also measurements in the strangeness sector will be feasible. After the submission of the Letter of Intent at the end of 2022, a technical proposal should be completed by 2024, with the plan of taking data from 2029. Together with the first measurements of open charm planned by NA61/SHINE in the immediate future, these two experiments will be able to advance our knowledge of the features of high-$\mu_{\rm B}$ QGP, complementing efforts that are or will be undertaken at other facilities (FAIR, RHIC, NICA, J-PARC).

\bibliographystyle{utphys}
\bibliography{biblio}


\end{document}